\documentclass[amstex,12pt]{article}
\usepackage{amsmath}
\usepackage{amsfonts}
\def\vk{\vec k}
\def\hxi{\hat\xi}

\def\he{\hat\eta}
\def\ve{\varepsilon}
\def\b{\beta}

\begin{document}

\begin{center}
{\bf{\Large Correlation functions in the Coulomb branch of \\
 \ \\
${\mathcal N}=4$ $SYM$ from AdS/CFT correspondence}}

\vspace*{1cm}

R.C.Rashkov\footnote{e-mails: rrachkov@sfu.ca; rash@phys.uni-sofia.bg,
on leave of absence from Dept. of Physics, Sofia University, 1164 Sofia,
Bulgaria}
and K.S.Viswanathan\footnote{e-mail: kviswana@sfu.ca}\\

\ \\
Department of Physics, Simon Fraser University \\
Burnaby, BC, V5A 1S6, Canada
\end{center}
\vspace*{.8cm}

\begin{abstract}

 We study $SU(N)\to S(U(N/2)\times U(N/2))$ symmetry breaking in
 ${\mathcal N}=4$ $SYM$ via AdS/CFT correspondence.  Two stacks of
 $\frac{N}{2}$ parallel $D3$ branes are separated by a distance $2
 {\bf{d}}$ in the transverse directions.  This leads to an interaction
 between different $l$-wave dilatonic KK modes when expanded in
 $S^{5}$ harmonics.  We calculate certain two point correlation
 functions of the dual gauge theory in the decoupling limit
 $\frac{d}{\alpha^{\prime}}\ll 1$ .  Due to mode mixing, the diagonal
 correlation functions have $\frac{1}{N}$ conformal-like correction as
 well as deformation terms.  The off-diagonal correlators are also
 nonvanishing and their leading order is $\frac{1}{N}$.  We discuss
 briefly the spectrum of the glueball exitations.
\end{abstract}

\section{Introduction}

A recent development in the understanding of the string theory was
a conjecture by
Maldacena  \cite{Malda} according to which, in some
appropriate
low energy limit, there is a duality between ${\mathcal N}=4$
supersymmetric
$SU(N)$ gauge theory and type {\bf II} strings on $AdS_{d+1}
\times{\mathcal M}$ background. This is nothing but a manifestation
of the
old ideas \cite{t'H},\cite{Pol} involving the deep connection between gauge
theories and strings. Therefore, from this point of view
it is clear that many problems in
large $N$ gauge theories can be studied  via supergravity in
AdS spaces\footnote{For a recent review see for instance \cite{Aha}.}.
The precise meaning and  the explicit rules for relating AdS supergravity
and the type {\bf IIB} string theory were given in \cite{GKP} and \cite{W}.
Namely, the explicit form of the correspondence between the correlation
functions of the operators in conformal field theory (CFT) living on the
asymptotic conformal boundary of the $d+1$ dimensional anti-de Sitter
space and type {\bf II} supergravity action evaluated on the classical
solutions $\Phi^i$ was proposed to be:
$$
\langle exp\int\limits_{S^d}\Phi^i_0{\mathcal O_i}\rangle_{CFT} =
exp\left(-S[\Phi^i]\right)
$$
where ${\mathcal O_i}$'s are conformal operators in CFT and $\Phi^i_0=
\Phi^i_{|\partial AdS}$ is the boundary value of the classical
solutions which serve as a source for the primary operators in CFT.
The proposed correspondence has been examined in a number of calculations
of the
correlation functions of operators in CFT  using the classical solutions
of various fields from the spectrum of AdS supergravity. These, even
incomplete proofs of the validity of the holographic principle, supply
us with a powerful tool for studying strongly coupled gauge theories. Most
of the intensive investigations of AdS/CFT correspondence
however are for the case of supergravity in pure
AdS background, i.e. at the origin of the moduli space where there is an
enhanced superconformal symmetry and many problems can be managed
exactly even at strong coupling.

Departure from the origin of the moduli space means allowing
scalar fields to have non-zero expectation values and therefore to move
to the Coulomb branch of the theory. The vacua in the Coulomb branch are
maximally supersymmetric but some of the gauge symmetry is broken by the
expectation value of scalar fields. These cases correspond to certain
solutions of the supergravity field equations in which the bulk space
geometry approaches AdS near the boundary, but differs from it in the
interior. From holographic correspondence  point of view  such geometries
are related either to a relevant deformation of the conformal field
theories, or to existence of states of the CFT with vacuum expectation
value which sets the energy scale.
In a recent paper by Klebanov and Witten \cite{KW} a more general
investigation of
the symmetry breaking is given and applied to several examples.
There are recent papers
studying the Coulomb branch of the gauge theories in four dimensions
\cite{{KLT},{FGPW},{BSf},{CR},{GR},{Fr}} which are based on continuous
spherically
symmetric distributions of large number of $D3$ branes. They, in fact, extend
the results of \cite{{BKL},{Sf}} by investigating different probes in the bulk
of the AdS space and their "holographic images" on the boundary. Since the
effective background is asymptotically $AdS_5\times S^5$ one can expand
the fields in spherical harmonics and solve the supergravity equations of
motion for the Kaluza-Klein (KK) modes. In this case, due to spherical
symmetry of the distribution of the $D3$ branes the different KK modes
decouple, thus enabling one to find  exact solutions of the field equations
using AdS/CFT correspondence to obtain the two point correlation functions.

In the present paper we will consider the simplest example of two
stacks of equal number
 coincident $D3$ branes separated by a distance $2\vec d$, which
corresponds to a vacuum state of the gauge theory where the gauge
$SU(N)$ symmetry is broken down to $S(U(N/2)\times U(N/2))$ by the
expectation value of scalar fields.

In Section 2 we consider the supergravity solutions in the presence of
two center $D3$ brane geometry  and the reduction of the 10
dimensional dilaton action to five dimensional one by integrating over
$S^{5}$.  It turns out that the equations of motion for the dilatonic KK
modes are coupled.  This coupling modifies the two-point correlation
functions, as discussed, in Section 3.  The diagonal correlation functions
of the operators ${\mathcal O}^n$ in the gauge theory are modified by a
$1/N$ correction term to the conformal-like part as a consequence  of the
coupling between the different modes.  In contrast to the conformal
case and the case of uniformly distributed $D3$-branes, the correlation
functions  $\langle{\mathcal O}^n(\vec x){\mathcal O}^{n+2}(\vec
y)\rangle$ are nonvanishing and are of order
$1/N$ in our approximation.  In addition, all correlators have a
deformation term which, after a Wick rotation, exhibits a pole
structure.  We interpret the appearance of double poles as a
degenerate spectrum of glueballs.  All these questions are subject of
consideration in Section 4.  We summarize, interpret and comment on
our results in the Concluding section.

\section{Two stacks of $D3$ branes: equations and solutions}

Let us consider the case of two center $D3$ branes .
In this case, a large number $N$ of $D3$ branes are separated into two
parallel stacks of $N/2$ branes each. Their separation distance introduces an
energy scale in the theory. The $SU(N)$ gauge symmetry is broken down to
$S(U(N/2)\times U(N/2))$.
>From the gauge theory  side this
corresponds to Higgsing of the gauge theory by scalar fields:
\begin{align}
\vec X=
\begin{pmatrix}
\vec d_1.I&0\\
0&\vec d_2.I
\end{pmatrix}
\label{2.1}
\end{align}
where $I$ is the $(N/2)\times(N/2)$ identity matrix and $\vec d_i$ is the
position of the i-th stack in the transverse direction.  Let us parametrize
the moduli space in terms of gauge invariant fields as in \cite{Rang,
W}:
\begin{equation}
{\mathcal O}^I=C^I_{i_1\dots i_k}Tr\left(X^{i_1}\dots X^{i_k}\right)
\label{co.1}
\end{equation}
where $ C^I_{i_1\dots i_k}$ is a totally symmetric traceless rank $k$,
$SO(6)$ tensor and $X^i$'s are as in (\ref{2.1}).  The Higgsing gives
expectation values for the chiral operators ${\mathcal O}^I$
\footnote{We call these operators chiral in the sense of \cite{Rang}}
which can be expressed in terms of order parameters $d_i$
\cite{{KW},{Rang}}:
$$
\langle{\mathcal O}^I\rangle \sim \prod\limits_{k=1}^n\left(d_1\right)_{
i_k}+
\prod\limits_{l=1}^n\left(d_2\right)_{i_l}-traces
$$

To describe  correlation functions on the Coulomb branch let us start
with the ten dimensional Type {\bf II} supergravity in the two-centered $D3$
brane background:
\begin{equation}
ds^2=H^{-\frac 12}\left(-dt+dx_1+dx_2+dx_3\right) +H^{\frac 12}
\sum\limits_{j=1}^6 dy_j^2
\label{2.2}
\end{equation}
where $H$ is the Green's function with two separated sources of equal
charge:
\begin{equation}
H=\frac{L^4}{2}\left(\frac{1}{|\vec y-\vec d_1|^4}+
\frac{1}{|\vec y-\vec d_2|^4}\right)
\label{2.3}
\end{equation}
In (\ref{2.3}), $\vec y$ are the coordinates normal  to the branes
\footnote{It is clear that at large $r$ the space is asymptotically
$AdS_5\times S^5$.  Since we consider the large $N$ case we have
neglected the one in $H$.  $L^4=4\pi g_s{\alpha'}^2N$.}.

Let us discuss briefly the geometry of this configuration and the decoupling
limit. Each of the two stacks develops a throat (geometry of which is AdS)
but they are located at different points of $S^5$, separated by a
distance $2\vec d$. In each throat the dilaton field can be expanded in
$S^{5}$ scalar spherical harmonics which will have non vanishing overlap
with the corresponding expansion around  the other throat.  For small $d$
(compared to the radius of AdS) and large $r$ the two throats meet and in
the limit $r\to\infty$ share a flat four dimensional boundary.  In this
case it is natural to make an appropriate shift in the, $\vec y$-direction,
choosing the origin so that $\vec d_1=-\vec d_2=\vec d$ and therefore:
\begin{equation}
H=\frac{L^4}{2}\left(\frac{1}{|\vec y+\vec d|^4}+
\frac{1}{|\vec y-\vec d|^4}\right)
\label{-}
\end{equation}
which allows us to consider the problem as effectively single
centered.  Due to  interaction between the different KK modes in the
bulk, it is expected that their images on the boundary result in a  mixing
between the correlation functions of operators of different conformal
dimensions.  We would like to point out that starting from Type IIB
string theory and taking the decoupling limit $\alpha'\to 0$ while
keeping $d/\alpha'$ fixed, we obtain an ${\mathcal N}=4$, 4-dimensional
$SYM$ theory with gauge group $SU(N)$ spontaneously broken down to
$S(U(N/2)\times U(N/2))$ by Higgs scalar expectation value in the
adjoint representation of $SU(N)$.  The Higgs scalars parametrize the
positions of the stacks.  We use the supergravity approximation which
is valid for $d<<L^2$.  In general one must distinguish between three
different regions: the first one is near the location of the branes.
Here the throats are well separated.  The second is where the throats
come together, but this region is still far from the boundary.  The
last region is near the boundary where the two throats effectively
coincide.  Since our coordinate system is symmetric with respect to
the positions of the stacks, the expansion over spherical harmonics
qualitatively changes on the boundary $r=|\vec d|$.  There we will
apply the matching condition technique in order to ensure well defined
smooth solutions in the bulk.  Therefore, using this choice of the
origin of the coordinate system one can expand the harmonic function
$H$ in Taylor series in the form:
\begin{equation}
H=\left(\frac {L}{r_>}\right)^4\left[1+\sum\limits_{l=1}^\infty
2^{2l}(2l+1)\left(\frac{r_<}{r_>}\right)^{2l}Y^{2l}(\Omega_5)\right]
\label{hg}
\end{equation}
where $Y^{2l}=\sum C^{2l}_{i_1\dots i_{2l}}y^{i_1}\dots y^{i_{2l}}$
 and $r_>,r_<$ are the larger or smaller of $r=|\vec y|$ and $d$ respectively.

 Let us now consider the kinetic part of the 10D dilaton action:
\begin{equation}
S^{\Phi}_{10}\sim\int d^{10}X\sqrt{G}G^{\mu\nu}\partial_\mu\Phi
\partial_\nu\Phi +\dots
\label{action}
\end{equation}
After integration by parts the above action takes the form:
\begin{equation}
S^{\Phi}_{10}=S^{kin}_{10}+S_B
\label{action1}
\end{equation}
where:
\begin{equation}
S^{kin}_{10}\sim\int\limits_{AdS}d^{10}X\left(H\Phi\nabla^2_{||}\Phi+
\Phi\nabla^2_{\perp}\Phi\right)
\label{kin}
\end{equation}
and
\begin{equation}
S_B\sim\int\limits_{\partial AdS}d\Sigma^\mu\Phi\nabla_\mu\Phi
\label{b}
\end{equation}
One can substitute for the dilaton field its expansion in $S^5$ scalar
spherical harmonics $Y^I(\Omega_5)$:
\begin{equation}
\Phi(\vec x,\vec y)=\sum\limits_{I=0}^\infty\Phi^I(\vec
x,r)Y^I(\Omega_5)
\label{harmo}
\end{equation}
where $r=|\vec y|$, and to obtain the five dimensional action in the
background metric (\ref{2.2}) we integrate over $S^5$ .  As it was noted in
a number of papers (see for instance \cite{KR}), we have an infinite
tower of KK modes $\Phi^I$ and there are no fundamental physical
reasons to truncate.  The resulting boundary action takes the form:
\begin{equation}
S_B=\int\limits_{r\to\infty}d^4xr^5\left[ A_>\Phi^0
\frac{\partial}{\partial r}\Phi^0+ \sum\limits_{I,J=1}^\infty A_>^{IJ}
\Phi^I\frac{\partial}{\partial r}\Phi^J + \sum\limits_{I=0}^\infty
A_>^I\frac{\partial}{\partial r}\left(\Phi^0\Phi^I\right)\right]
\label{sb}
\end{equation}
where\footnote{We will use the notations and normalization of the
spherical harmonics as in \cite{Rang}.}:
\begin{equation}
A_>=\int d\Omega_5 H_>=V_5
\end{equation}
\begin{equation}
A^I_>=\int d\Omega_5 H_>Y^I=
V_5\left(\frac
Lr\right)^4\frac{1}{l+1}\left(\frac dr\right)^{2l}
\delta^{I,2l}
\end{equation}
\begin{align}
A^{IJ}_>=&\int d\Omega_5 H_>Y^IY^J=
V_5\left(\frac Lr\right)^4
\frac{\delta^{IJ}}
{2^{I-1}(I+1)(I+2)}
\notag \\
& +V_5\frac{L^4}{r^4}\sum\limits_{l=1}^\infty
2^{2l}(2l+1)\left(\frac dr\right)^{2l}
a(2l,I,J)\langle C^{2l}C^IC^J\rangle
\end{align}
 As in \cite{Rang} we use the following normalization:
\begin{align}
& a(L,I,J)=\frac{1}{(\Sigma+2)!2^{\Sigma-1}}.
\frac{k_L!k_I!k_J!}{\alpha_1!\alpha_2!\alpha_3!}
\notag \\
&\Sigma=\frac 12(k_L+k_I+k_J);\quad
\alpha_1=\Sigma-k_L;\quad\alpha_2=\Sigma-k_I;
\quad\alpha_3=\Sigma-k_J
\label{norm}
\end{align}
We will see that the last term in (\ref{sb}) produces only local terms on
the  boundary and therefore is irrelevant for our considerations.

The kinetic term can be easily reduced to five dimensional effective
action in the form \cite{CR} (from now on we will denote $\Phi^0$ by
$\Phi$):
\begin{align}
S^{kin}_\Phi=&\int d^4xdr\,\,r^5\left\{
\Phi\left[\left(\int d\Omega_5 H\right)\nabla^2_{||}
+V_5\frac{1}{r^5}\partial_r r^5\partial_r\right]\Phi
\right. \notag \\
&  +\left(\int d\Omega_5 HY^I\right) \left[\Phi\nabla^2_{||}\Phi^I +\Phi^I
\nabla^2_{||}\Phi\right]+
\left(\int d\Omega_5 HY^IY^J\right)\Phi^I\nabla^2_{||}\Phi^J
\notag \\
&  + \left.\left(\int d\Omega_5 Y^IY^J\right)\Phi^I
\frac{1}{r^5}\partial_r r^5\partial_r\Phi^J+\frac{1}{r^2}
\left(\int d\Omega_5 Y^I\nabla^2_{S^5}Y^J\right)\Phi^I\Phi^J
\right\}
\label{5kin}
\end{align}
It is straightforward to derive the equations of motion following from
(\ref{5kin}) in the  form:
\begin{equation}
V_5\triangle_r\Phi+
\left(\int d\Omega_5H\right)\nabla^2_{||}\Phi+
\left(\int d\Omega_5Y^I\right)\nabla^2_{||}\Phi^I=0
\label{eqm0}
\end{equation}
\begin{align}
\left(\int d\Omega_5Y^IY^J\right)
\triangle_r\Phi^J+
&
\frac{1}{r^2}\left(\int d\Omega_5Y^I\nabla_{S^5}^2Y^J\right)\Phi^J+
\notag \\
&\left(\int d\Omega_5HY^IY^J\right)\nabla^2_{||}\Phi^J +
\left(\int d\Omega_5HY^I\right)\nabla^2_{||}\Phi=0,
\label{eqm}
\end{align}
where $\triangle_r=(1/r^5)\partial_r(r^5\partial_r)$.

Since the metric has different forms in the regions of $r<d$ and $r>d$
we will have two systems of equations - one for each region.  Taking
into account the expression for $H$ (\ref{hg}) and the orthogonality
of the spherical harmonics (with normalization as in \cite{Rang}) the
equations of motion are given by:

a) $r>d$ region

\begin{equation}
\triangle_r\Phi(r,\vk)-\frac{L^4}{r^4}k^2\Phi(r,\vk)=
\frac{L^4}{r^4}k^2\sum\limits_{l=1}^\infty\frac{1}{l+1}
\left(\frac dr\right)^{2l}\Phi^{2l}(r,\vk) .
\label{eq>0}
\end{equation}
\begin{align}
&\triangle_r\Phi^n(r,\vk)-\frac{n(n+4)}{r^2}\Phi^n(r,\vk)
-\frac{L^4}{r^4}k^2\Phi^n(r,\vk)=
\notag \\
& \frac{L^4}{r^4}k^22^{n-1}(n+1)(n+2)\sum\limits_{l=1}^\infty
2^{2l}(2l+1)\left(\frac dr\right)^{2l}
\sum\limits_{m=|2l-n|}^{2l+n}A^{2l,n,m}\Phi^m(r,\vk)
\notag \\
&\hspace*{5cm}
+\frac{L^4}{r^4}k^22^{2l}(2l+1)\left(\frac dr\right)^{2l}\delta^{2l,n}
\Phi(r,\vk) .
\label{eq>n}
\end{align}

b) $r<d$ region

\begin{equation}
\triangle_r\Phi(r,\vk)-\frac{L^4}{d^4}k^2\Phi(r,\vk)=
\frac{L^4}{d^4}k^2\sum\limits_{l=1}^\infty\frac{1}{l+1}
\left(\frac rd\right)^{2l}\Phi^{2l}(r,\vk)
\label{eq<0}
\end{equation}
\begin{align}
&\triangle_r\Phi^n(r,\vk)-\frac{n(n+4)}{r^2}\Phi^n(r,\vk)
-\frac{L^4}{d^4}k^2\Phi^n(r,\vk)=
\notag \\
& \frac{L^4}{d^4}k^22^{n-1}(n+1)(n+2)\sum\limits_{l=1}^\infty
2^{2l}(2l+1)\left(\frac rd\right)^{2l}
\sum\limits_{m=|2l-n|}^{2l+n}A^{2l,n,m}\Phi^m(r,\vk)
\notag \\
&\hspace*{5cm}
+\frac{L^4}{d^4}k^22^{2l}(2l+1)\left(\frac rd\right)^{2l}\delta^{2l,n}
\Phi(r,\vk)
\label{eq<n}
\end{align}
In the above expressions we used the following notation:
\begin{equation}
 A^{L,I,J}=a(L,I,J)\langle C^LC^IC^J\rangle
\end{equation}
and $a(L,I,J)$ and $<C^LC^IC^J>$ are as in eq.(\ref{norm}).
We see that the equation for the different KK modes of the
dilaton field $\Phi^n$ are coupled.
In general, according to the AdS/CFT correspondence, the operators
${\mathcal O}^I$ are mapped to conformal fluctuations in the metrics of
$AdS_5$ and $S^5$. In our case this implies that the dilaton action
projected on the boundary will produce correlation functions of the
operators ${\mathcal O}^I\sim tr\,(F^2X^I)$.
In contrast to the uniformly distributed $D3$ branes case, there
is coupling between the different KK modes, which signals a
nontrivial contribution to the corresponding correlations
functions.
The mixing between the KK modes
means that, due to the fact that we have a broken conformal symmetry, the
sources coming from gravity solutions will produce in general
mixed correlation functions. We will see how it happens in the next
section.

The field equations in the two regions can be transformed into
Bessel equations by the following substitutions:

a) for $r>d$ let us define:
\begin{align}
&\Phi^n(r,\vk)=\xi^2f_n(\xi,\vk);
&\xi=\frac dr; \qquad
&\hat\xi=\kappa\xi
\notag \\
&\hat\xi=\kappa\xi;
&\kappa^2=\frac{L^4}{d^2}k^2;\qquad
&(\xi\leq 1)
\notag
\end{align}
then the equations (\ref{eq>0}, \ref{eq>n}) become:
\begin{equation}
f''(\hxi)+\frac{1}{\hxi}f'(\hxi)-
\left(1+\frac{4}{\hxi^2}\right)f(\hxi)=
\sum\limits_{l=1}^\infty\frac{1}{(l+1)\kappa^{2l}}\hxi^{2l}f_{2l}(\hxi)
\label{feq>0}
\end{equation}
\begin{align}
f_n''(\hxi)&+\frac{1}{\hxi}f_n'(\hxi)-
\left(1+\frac{(n+2)^2}{\hxi^2}\right)f_n(\hxi)
\notag \\
&=2^{n-1}(n+1)(n+2)\sum\limits_{l=1}^\infty
\frac{2^{2l}(2l+1)}{\kappa^{2l}}\hxi^{2l}
\sum\limits_{m=|2l-n|}^{2l+n}A^{2l,n,m}f_m(\hxi)
\notag \\
&\hspace*{5cm}
+\frac{2^{2l}(2l+1)}{\kappa^{2l}}\hxi^{2l}\delta^{2l,n}f(\hxi) .
\label{feq>n}
\end{align}

b) For $r<d$, defining:
\begin{align}
&\Phi^n(r,\vk)=\eta^{-2}f_n(\eta,\vk);
&\eta=\frac rd;\qquad
&\hat\eta=\kappa\eta
\notag \\
&\hat\eta=\kappa\eta;
&\kappa^2=\frac{L^4}{d^2}k^2;\qquad
&(\eta\leq 1)
\notag
\end{align}
the equations (\ref{eq<0}, \ref{eq<n}) become:
\begin{equation}
f''(\he)+\frac{1}{\he}f'(\he)-
\left(1+\frac{4}{\hxi^2}\right)f(\he)=
\sum\limits_{l=1}^\infty\frac{1}{(l+1)\kappa^{2l}}\he^{2l}f_{2l}(\he)
\label{feq<0}
\end{equation}
\begin{align}
f_n''(\he)&+\frac{1}{\he}f_n'(\he)-
\left(1+\frac{(n+2)^2}{\he^2}\right)f_n(\he)
\notag \\
&=2^{n-1}(n+1)(n+2)\sum\limits_{l=1}^\infty
\frac{2^{2l}(2l+1)}{\kappa^{2l}}\he^{2l}
\sum\limits_{m=|2l-n|}^{2l+n}A^{2l,n,m}f_m(\he)
\notag \\
&\hspace*{5cm}
+\frac{2^{2l}(2l+1)}{\kappa^{2l}}\he^{2l}\delta^{2l,n}f(\he) .
\label{feq<n}
\end{align}
where primes on $f$ denote differentiation with respect to its
argument.  The terms on the right hand side of the equations above
involve coupling to KK modes.   For small
$d$ and large $N$, we can solve these equations perturbatively.  One can
see that the right hand side of both systems of equations is readily
expanded in $d^2/L^4=\pi^2m_W^2\frac{{\alpha'}^2}{L^4}$, which in the
decoupling limit is small. Here, $m_{W}=\frac{d}{\pi \alpha^{\prime}}$
is the $W$ boson mass, generated by symmetry breaking. This is precisely the
region of validity of our approximation, i.e.  for energies below the
$W$-boson mass.  In the zeroth order we ignore the coupling of the
modes and these solutions are used as inhomogeneous terms in obtaining
the first order solution.

Let us consider the zeroth order approximation. The solutions of
the homogeneous system for $r>d$ are:
\begin{subequations}
\begin{align}
f_n(\hxi,\vk)=\begin{cases}
&I_{n+2}(\hxi)\cr
&K_{n+2}(\hxi)
\end{cases}
\quad n=0,1,\dots
\label{sf>n}
\end{align}
\end{subequations}
 and therefore the general solution for the dilatonic
modes is:
\begin{equation}
\Phi^n(\kappa\xi)=\xi^2\left[K_{n+2}(\kappa\xi)\hat\mu^{(0)}_n(\vk)
+I_{n+2}(\kappa\xi)\mu^{(0)}_n(\vk)\right]
\label{sF>n}
\end{equation}
where $\hat\mu^{(0)}_n(\vk)$ and $\mu^{(0)}_n(\vk)$ are yet to be determined
coefficients.

The solutions in $r<d$ region are:
\begin{align}
f_n(\he,\vk)=\begin{cases}
&I_{n+2}(\kappa\he)\cr
&K_{n+2}(\kappa\he)
\end{cases}
\quad n=0,1,\dots
\label{sf<n}
\end{align}
or, taking into account that $\eta=1/\xi$:
\begin{equation}
\Phi^n(\frac\kappa\xi)=\xi^2\left[K_{n+2}(\frac\kappa\xi)
\hat\alpha^{(0)}_n(\vk)
+I_{n+2}(\frac\kappa\xi)\alpha^{(0)}_n(\vk)\right]
\label{F<n}
\end{equation}
In general, our solution must be regular in the interior of AdS space so
that the solutions for $r>d$ must be extended in $r<d$ region to a
regular solution. The only regular solution in this region is:
\begin{equation}
\Phi^n(\frac\kappa\xi)=\xi^2I_{n+2}(\frac\kappa\xi)\alpha^{(0)}_n (\vk)
\label{sF<n}
\end{equation}
In addition, the two solutions (\ref{sF>n}) and (\ref{sF<n}) must be
matched on the boundary of the two regions ($\xi=1$):
\begin{align}
&\Phi^n\left(\frac\kappa\xi\right)_{|\xi=1}=
\Phi^n(\kappa\xi)_{|\xi=1}
\label{match1} \\
&\left[\Phi^n\left(\frac\kappa\xi\right)\right]'_{|\xi=1}=
\left[\Phi^n(\kappa\xi)\right]'_{|\xi=1}
\label{match2}
\end{align}
These conditions uniquely determine the coefficients $\mu^{(0)}_n$ and
$\alpha^{(0)}_n$:
\begin{equation}
\mu^{(0)}_n(\vk)= -\frac{
K_{n+2}(\kappa)I_{n+2}'(\kappa)+K_{n+2}'(\kappa)I_{n+2}(\kappa)}
{2I_{n+2}(\kappa)I_{n+2}'(\kappa)}
\end{equation}
\begin{equation}
\alpha^{(0)}_n(\vk)= \frac{
K_{n+2}(\kappa)I_{n+2}'(\kappa)-K_{n+2}'(\kappa)I_{n+2}(\kappa)}
{2I_{n+2}(\kappa)I_{n+2}'(\kappa)}
\end{equation}
To determine the last unknown coefficient $\hat\mu^{(0)}_n$, note that
the scalar field near the boundary behaves as
\cite{{GKP},{W}}:
$$
Q(\xi,\vk)\underset{\xi\to 0}{\approx}
\xi^{\frac d2 -\nu}Q_b(\vk)+ \dots ,
$$
where $Q_b(\vk)$ is the boundary value of $Q$ and $\nu$ is the spectral
parameter in the equations of motion.
Applying the above formula to the n-$th$ dilatonic KK mode and using the
behaviour of the Bessel functions of small argument, one can see that the
only contribution to the above behaviour comes from $K_{n+2}$ and
therefore:
$$
\Phi^n_b(\vk)=\frac{(n+1)!}{2}\left(\frac{\kappa}{2}\right)^{-(n+2)}
\hat\mu^{(0)}_n(\vk)
$$
or:
\begin{equation}
\hat\mu^{(0)}_n(\vk)=\frac{2}{(n+1)!}
\left(\frac{\kappa}{2}\right)^{n+2}
\Phi^n_b(\vk)
\end{equation}
With this, all the coefficients in the zeroth order solutions are
completely determined. We subsitute the zeroth order solutions on the right
hand side of (\ref{feq>0},\ref{feq>n})and (\ref{feq<0},\ref{feq<n}) to
find the first order correction.
The equations for the first order ($l=1$) are as follows:

a) $r>d$ region
\begin{equation}
f''(\hxi)+\frac{1}{\hxi}f'(\hxi)-
\left(1+\frac{4}{\hxi^2}\right)f(\hxi)=
\frac{1}{2\kappa^2}\hxi^2\left[K_4(\hxi)\hat\mu^{(0)}_2(\vk)
+I_4(\hxi)\mu^{(0)}_2(\vk)\right]
\label{1fe>0}
\end{equation}
\begin{align}
f_n''(\hxi)&+\frac{1}{\hxi}f_n'(\hxi)-
\left(1+\frac{(n+2)^2}{\hxi^2}\right)f_n(\hxi)
\notag \\
& =3\frac{2^{n+1}(n+1)(n+2)}{\kappa^2}\hxi^2\left[
A^{2,n,n-2}\left(K_n(\hxi)\hat\mu^{(0)}_{n-2}(\vk)
+I_n(\hxi)\mu^{(0)}_{n-2}(\vk)\right)\right.
\notag \\
& \hspace*{3.5cm}+A^{2,n,n}\left(K_{n+2}(\hxi)\hat\mu^{(0)}_{n}(\vk)
+I_{n+2}(\hxi)\mu^{(0)}_{n}(\vk)\right)
\notag \\
& \hspace*{3.5cm}\left.+A^{2,n,n+2}\left(K_{n+4}(\hxi)\hat\mu^{(0)}_{n+2}(\vk)
+I_{n+4}(\hxi)\mu^{(0)}_{n+2}(\vk)\right)\right]
\label{1fe>n}
\end{align}

b)  $r<d$ region
\begin{equation}
f''(\he)+\frac{1}{\he}f'(\he)-
\left(1+\frac{4}{\he^2}\right)f(\he)=
\frac{1}{2\kappa^2}\he^2I_4(\he)\alpha^{(0)}_2(\vk)
\label{1fe<0}
\end{equation}
\begin{align}
f_n''(\he)&+\frac{1}{\he}f_n'(\he)-
\left(1+\frac{(n+2)^2}{\he^2}\right)f_n(\he)
\notag \\
& =3\frac{2^{n+1}(n+1)(n+2)}{\kappa^2}\he^2\left[
A^{2,n,n-2}I_n(\he)\alpha^{(0)}_{n-2}(\vk)\right.
\notag \\
& \hspace*{3cm}\left.+A^{2,n,n}I_{n+2}(\he)\alpha^{(0)}_{n}(\vk)
+A^{2,n,n+2}
I_{n+4}(\he)\alpha^{(0)}_{n+2}(\vk)\right]
\label{1fe<n}
\end{align}
Solving these equations we find that the general solutions for $r>d$
are\footnote{See Appendix for details.}:
\begin{align}
f_n(\hxi)=&K_{n+2}(\hxi)\hat\mu^{(1)}_n(\vk)
+I_{n+2}(\hxi)\mu^{(1)}_n(\vk)
\notag \\
&+\frac{2^{n}(n+1)(n+2)}{\kappa^2}
\left\{
A^{2,n,n-2}\left(\hxi^3I_{n-1}(\hxi)\mu^{(0)}_{n-2}-
\hxi^3K_{n-1}(\hxi)\hat\mu^{(0)}_{n-2}\right)\right.
\notag \\
&\hspace*{2.5cm}
+A^{2,n,n}\left[
\left(\hxi^3I_{n-1}(\hxi)+3(n+1)\hxi^2I_n(\hxi)\right)\mu^{(0)}_n
\right.
\notag \\
&\hspace*{2.5cm}
\left.
-\left(\hxi^3K_{n-1}(\hxi)+3(n+1)\hxi^2K_n(\hxi)\right)\hat\mu^{(0)}_n
\right]
\notag \\
&\hspace*{2.5cm}
\left.
+A^{2,n,n+2}\left(\hxi^3I_{n+5}(\hxi)\mu^{(0)}_{n+2}-
\hxi^3K_{n+5}(\hxi)\hat\mu^{(0)}_{n+2}\right)\right\}
\label{1sf>n}
\end{align}
\begin{equation}
f(\hxi)=K_2(\hxi)\hat\mu^{(1)}(\vk)+I_2(\hxi)\mu^{(1)}(\vk)+
\frac{1}{12\kappa^2}\left(\hxi^3I_5(\hxi)\mu^{(0)}_2-
\hxi^3K_5(\hxi)\hat\mu^{(0)}_2\right)
\label{1sf>0}
\end{equation}
For the region $r<d$, the regular solutions are:
\begin{align}
f_n(\he)=&I_{n+2}(\he)\alpha^{(1)}_n+
\frac{2^{n+1}(n+1)(n+2)}{\kappa^2}
\left[A^{2,n,n-2}\he^3I_{n-1}(\he)\alpha^{(0)}_{n-2}
\right. \notag \\
&
\left. +A^{2,n,n}
\left(\he^3I_{n-1}(\he)+3(n+1)\he^2I_n(\he)\right)\alpha^{(0)}_n
+A^{2,n,n+2}\he^3I_{n+5}(\he)\alpha^{(0)}_{n+2}\right]
\label{1sf<n}
\end{align}
\begin{equation}
f(\he)=I_2(\he)\alpha^{(1)}(\vk)+
\frac{1}{12\kappa^2}\he^3I_5(\hxi)\alpha^{(0)}_2
\label{1sf<0}
\end{equation}
Once again we will use the required behaviour of the solutions near the
boundary. Expanding the Bessel functions in (\ref{1sf>n}) and
(\ref{1sf>0}) for small argument we find that:
\begin{equation}
\hat\mu^{(1)}_n(\kappa)=
\frac{2}{(n+1)!}\left[\Phi^{n}_b(\vk)+
2^{n+1}\frac{(n+1)!}{n!}A^{2,n,n+2}\Phi^{n+2}_b(\vk)\right]
\left(\frac{\kappa}{2}\right)^{n+2}
\end{equation}
\begin{equation}
\hat\mu^{(1)}_0(\kappa)=
2\left[\Phi^{0}_b(\vk)+
\frac{2}{3}\Phi^{2}_b(\vk)\right]
\left(\frac{\kappa}{2}\right)^{2}
\end{equation}
As for the zeroth order solutions, the
coefficients $\hat\mu^{(1)}_n(\kappa)$ are determined by the boundary
data.  There is, however, an important difference - these coefficients
are determined by a linear combination of $\Phi^{n}_b(\vk)$ and
$\Phi^{n+2}_b(\vk)$.  The solutions for n-$th$ dilatonic KK mode
contain two as yet undetermined coefficients, but they can be
determined by the matching conditions on the boundary of the two
regions.  We will see later that these coefficients will play an
important role in interpreting the correlation functions and therefore
on the physics of the Coulomb branch.

To conclude this section let us summarize  the results we have
found.

Starting from 10D dilaton action in 2 stacks $D3$ brane background we
derived the coupled equations of motion for the KK harmonics.  We have
obtained explicit solutions to first order in $\frac{1}{\kappa^{2}}$.
The arbitrary coefficients in the solutions were determined by applying
the boundary conditions at $r=d$ as well as from the appropriate
asymptotic properties.  The most interesting results are contained in
eqns(47) and (48) which reflect the coupling of the $n$ and $n+2$, KK modes.
This leads to non conformal correction terms to the correlation functions.

\section{Correlation functions}

As mentioned in the Introduction  two stacks of parallel, separated
$D3$ branes preserve supersymmetry, since the Poincare supersymmetries
of the gauge theory are maintained but superconformal invariance is
broken by the Higgsing.  In the case under consideration, the $SU(N)$
symmetry is broken down to $S(U(N/2)\times U(N/2))$.  This will modify
the correlation functions which we study in this section.

>From the gauge theory side, according to AdS/CFT
correspondence, evaluation of the action on the classical solutions will
produce the correlation functions of the operators
$tr\left(F^2X^I\right)$ (we use the notations of \cite{Rang}:
$X^I=C^I_{i_1\dots i_k}X^{i_1}\dots X^{i_k}$).  As shown in the
Introduction, the nontrivial contributions will come from the terms of
the following form\footnote{One can easily check that the terms which
survive in the limit $\xi\to 0$ coming from
$A_I\Phi^0\partial_\xi\Phi^I$ contain only positive integer powers of
$k$ and hence are contact terms.} :
\begin{equation}
S_B=-\frac 12\lim\limits_{\ve\to 0}\int d^4x\ve^{-3}
\Phi^I\partial_\xi\Phi^I_{|\xi=\ve}
\label{cf}
\end{equation}

We saw in the previous Section that there are two undetermined
coefficients $\mu^{(1)}_n(\kappa)$ and $\alpha^{(1)}_n(\kappa)$ for every
$n$. In order to proceed with the derivation of the correlation
functions in the gauge theory we first determine these.
For this purpose, we have to impose on the first order solutions the
matching conditions (\ref{match1},\ref{match2})
on the boundary of the two regions ($\xi=1$)
(note that these are imposed now on the first order solutions). After
tedious but straightforward calculations we find the following
expression for $\mu^{(1)}_n(\kappa)$:
\begin{equation}
\mu^{(1)}_n(\kappa)=\b_{n,n-2}(\kappa)\Phi^{n-2}_b(\vk)+
\b_{n,n}(\kappa)\Phi^{n}_b(\vk)+
\b_{n,n+2}(\kappa)\Phi^{n+2}_b(\vk)
\label{mgen}
\end{equation}
where:
\begin{equation}
\b_{n,n}(\kappa)=\b^{(1)}_{n,n}(\kappa)+\b^{(2)}_{n,n}(\kappa)
\label{bgen}
\end{equation}
\begin{align}
\b^{(1)}_{n,n}=&\left\{-\left(K_{n+2}I_{n+2}'+K_{n+2}'I_{n+2}\right)
\right.
\notag \\
& +2^n(n+1)(n+2)A^{2,n,n}\left[\left(\kappa K_{n-1}+3(n+1)K_{n}
\right)I_{n+2}' \right.
\notag \\
&\left.\left.+\left( 3K_{n-1}+\kappa K_{n-1}'
+\frac{6(n+1)}{\kappa^2}K_{n}+
3(n+1)K_{n}'\right)I_{n+2}\right]\right\}\times
\notag \\
&\hspace*{6cm}\times\frac{1}{(n+1)!I_{n+2}I_{n+2}'}
\label{b1}
\end{align}
\begin{align}
\b^{(2)}_{n,n}=&
\frac{2^n(n+1)(n+2)A^{2,n,n}}{(n+1)!}
\left[\left(\kappa I_{n-1}+3(n+1)I_n\right){I_{n+2}'}^2
\right.
\notag \\
&\left. +\left(3I_{n-1}+\kappa I_{n-1}'+\frac{6(n+1)}{\kappa}I_{n}
+3(n+1)I_{n}'\right)I_{n+2}^2
\right]\frac{1}{I_{n+2}^2{I_{n+2}'}^2}
\label{b2}
\end{align}
\begin{equation}
\b_{n,n+2}(\kappa)=\b^{(1)}_{n,n+2}(\kappa)+\b^{(2)}_{n,n+2}(\kappa)
\label{bmixg}
\end{equation}
\begin{align}
\b^{(1)}_{n,n+2}=&
\frac{2^n(n+1)(n+2)A^{2,n,n+2}}{(n+1)!}
\left[
-2(n+4)\left(K_{n+2}I_{n+2}'+K_{n+2}'I_{n+2}\right)
\left(\frac{\kappa}{2}\right)^{-2}\right.
\notag \\
& \left.
\frac{K_{n+5}I_{n+2}'+\left(3K_{n+5}+\kappa K_{n+5}'\right) I_{n+2}}
{(n+2)(n+3)}\right]
\frac{1}{I_{n+2}I_{n+2}'}
\label{bmix1g}
\end{align}
\begin{align}
\b^{(2)}_{n,n+2}=&
\frac{2^n(n+1)(n+2)A^{2,n,n+2}}{2(n+3)!}
\left[\kappa I_{n+5}I_{n+2}'K_{n+4}I_{n+4}'
\right.
\notag \\
&\left. +\left( 3I_{n+5}+\kappa I_{n+5}'\right)
I_{n+2}K_{n+4}'I_{n+4}\right]
\frac{1}{I_{n+2}I_{n+2}'I_{n+4}I_{n+4}'}
\label{bmix2g}
\end{align}
\begin{equation}
\b_{n,n-2}(\kappa)=\b^{(1)}_{n,n-2}(\kappa)+\b^{(2)}_{n,n-2}(\kappa)
\label{bmixl}
\end{equation}
\begin{align}
\b^{(1)}_{n,n-2}=&
\frac{2^n(n+1)(n+2)A^{2,n,n-2}}{(n-1)!}\times
\notag \\
&\times
\left[
\left(3K_{n-1}+\kappa K_{n-1}'\right) I_{n+2}
+\kappa K_{n-1}I_{n+2}'\right]
\frac{1}{I_{n+2}I_{n+2}'}
\label{bmix1l}
\end{align}
\begin{align}
\b^{(2)}_{n,n-2}=&
\frac{2^n(n+1)(n+2)A^{2,n,n-2}}{2(n-1)!}
\left[\kappa I_{n-1}I_{n+2}'K_{n}I_{n}'
\right.
\notag \\
&\left. +\left( 3I_{n-2}+\kappa I_{n-2}'\right)
I_{n+2}K_{n}'I_{n}\right]
\frac{1}{I_{n+2}I_{n+2}'I_{n}I_{n}'}
\label{bmix2l}
\end{align}

Having completely determined these coefficients, we
simply have to subsitute
the explicit form of our first order solutions in the boundary term
action and  take the limit $\ve\to 0$. To see which terms survive this
limit, we must use the expansion of the Bessel functions for small
argument. After some lengthy but straightforward calculations we find the
following expression for the diagonal correlation functions in momentum
space ( we have omitted  terms of even integer powers of $k$ since
they produce in the position space $\delta$-function and its
derivatives):
\begin{align}
\langle{\mathcal O}^n(k){\mathcal O}^n(k')\rangle=&
-\delta(\vk+\vk')\left[
(-1)^n\frac{2^3}{(n+1)!(n+2)!}\left(\frac{\kappa}{2}\right)^{2n+4}
\,ln\,k \right. \notag \\
& +(-1)^n2^{n+1}\frac{4(n+1)^2(n+2)(n+3)}{(n+1)!^2}A^{2,n,n}
\left(\frac{\kappa}{2}\right)^{2n+2}\,ln\,k
\notag \\
&-(-1)^n2^{2n+2}\frac{(n+2)(5n^2+13n+12)}{(n-1)!(n-2)!}
\left(A^{2,n,n}\right)^2\left(\frac{\kappa}{2}\right)^{2n}\,ln\,k
\notag \\
& \hspace*{4cm}
\left.+D_{n,n}(\kappa)\right]
\label{dcfm}
\end{align}
Since we have mixing between the different KK dilatonic modes, the
off-diagonal correlation functions are also nonvanishing. We find for
them the following expression:
\begin{align}
\langle{\mathcal O}^n(k){\mathcal O}^{n+2}(k')\rangle=&
-\delta(\vk+\vk')\left[
(-1)^n2^{n+4}\frac{(n+2)(n+4)}{(n+1)!^2}A^{2,n,n+2}
\left(\frac{\kappa}{2}\right)^{2n+4}
\,ln\,k \right. \notag \\
& +(-1)^n2^{2n+4}\frac{(n+1)(n+2)^2(n+2)(n+3)(n+4)}{(n)!^2}
\times\notag \\
& \left.\times
A^{2,n,n}A^{2,n,n+2}
\left(\frac{\kappa}{2}\right)^{2n+2}\,ln\,k
+D_{n,n+2}(\kappa)\right]
\label{dcfm1}
\end{align}
In addition to  the conformal-like parts (integer powers of $\kappa$ times
$ln\,k$) the correlation functions contain deformations coming from
the coefficient functions $\mu^{(1)}_n$ entering the solutions.
Substituting the expressions for $\mu^{(1)}_n$ into the boundary
action and keeping the terms that survive $\xi\to 0$ limit one can
obtain the deformation functions $D_{n,m}$.  To simplify the final
expressions we introduce the following notations:
\begin{align}
&a_n(\kappa)=\frac{2}{(n+1)!}\left(\frac{\kappa}{2}\right)^{n+2};
\qquad
\b_m(\kappa)=-\frac{K_{n+2}I_{n+2}'+K_{n+2}'I_{n+2}}{2I_{n+2}I_{n+2}'}
\notag \\
& b_n(\kappa)=2^{n+2}\frac{(n+2)(n+4)}{n!}A^{2,n,n+2}
\notag
\end{align}
In these notations, we find for $D_{n,m}$ the expressions:
\begin{align}
&D_{n,n}(\kappa)=
\frac{2}{n+2}a_n\b_{n,n}\left(\frac{\kappa}{2}\right)^{n+2}
\notag \\
&-2^{n-2}(n-1)(n+1)(n+2)(5n+2)A^{2,n-2,n}a_n\b_n
\left(\frac{\kappa}{2}\right)^{n}\b_{n-2,n}
\notag \\
&+2^{n+1}(n+1)^2(n+2)(5n+3)A^{2,n,n}a_n^2\b_n
\left(\frac{\kappa}{2}\right)^{-2}+b_{n-2}\b_{n-2,n}
\left(\frac{\kappa}{2}\right)^{n+2}
\label{defd}
\end{align}
\begin{align}
&D_{n,n+2}(\kappa)=
\frac{2}{n+2}\left[a_n\b_{n,n+2}\left(\frac{\kappa}{2}\right)^{n+4}
+b_n\b_{n,n}\left(\frac{\kappa}{2}\right)^{n+2}\right]
\notag \\
&-2^{n}(n+1)(n+3)(n+4)(5n+12)A^{2,n,n+2}a_{n+2}\b_{n+2}
\b_{n,n}\left(\frac{\kappa}{2}\right)^{n+2}
\notag \\
&+ 2^{n+1}(n+1)^2(n+2)(5n+3)A^{2,n,n}
\left(\frac{\kappa}{2}\right)^{-2}a_{n+2}b_n\b_{n}
\notag \\
&-
2^{2n+2}(n+1)^3(n+2)^3(n+3)(n+4)(5n+3)A^{2,n,n}A^{2,n,n+2}
\left(\frac{\kappa}{2}\right)^{-4}a_{n}a_{n+2}\b_{n}
\label{defn}
\end{align}

Fourier transforming to position space we obtain the following for the
two point correlation functions:
\begin{align}
\frac{
\langle{\mathcal O}^n(\vec x){\mathcal O}^{n}(\vec y)\rangle}
{\left(\frac{L^2}{d}\right)^{2n+4}}=&
\frac{4(n+2)(n+3)}{\pi^2}.
\frac{1}{|\vec x-\vec y|^{2n+8}}
\notag \\
&+2^{n+1}\frac{d^2}{4\pi g_s{\alpha'}^2N}.
\frac{(n+1)^2(n+2)^2(n+3)}{\pi^2}.
\frac{1}{|\vec x-\vec y|^{2n+6}}
\notag \\
&\hspace*{3cm}+\hat D_{n,n}(\frac{d}{L^2}|\vec x-\vec y|)+O(1/N^2)
\label{dcfx}
\end{align}
\begin{align}
\frac{
\langle{\mathcal O}^n(\vec x){\mathcal O}^{n+2}(\vec y)\rangle}
{\left(\frac{L^2}{d}\right)^{2n+6}}=&
2^{n+3}\frac{d^2}{4\pi g_s{\alpha'}^2N}.
\frac{(n+1)^2(n+2)^2(n+3)(n+4)}{\pi^2}\times
\notag \\
&\times
A^{2,n,n+2}
\frac{1}{|\vec x-\vec y|^{2n+8}}+
\hat D_{n,n+2}(\frac{d}{L^2}|\vec x-\vec y|)+O(1/N^2)
\label{offcfx}
\end{align}
In  above, we have used  $L^4=4\pi g_s{\alpha'}^2N$, and the deformation
of the conformal part $\hat D_{n,m}(\frac{d}{L^2}|\vec x-\vec y|)$
is the Fourier transform of $D_{n,m}(\kappa)$. It is well known
that ${\mathcal O}^n\sim tr(F^2X^n)$. In contrast to the case of
spherically
symmetric, continuously distributed $D3$-branes, the mixed correlators are
nonvanishing. Keeping the first order terms in the solutions we
have found
that the correlation functions contain $\frac{1}{N}$ corrections. The
leading term
in the mixed correlators, which comes from the first order solutions, is
proportional to $1/N$. One can see directly from the explicit form of
the correlation functions that in the conformal limit $d\to 0$ (or
equivalently $<X>\to 0$) the $\frac{1}{N}$ correction term vanishes (the
deformations $D_{n,m}$ also vanish).  In that limit the mixed
correlation functions also vanish and the conformal symmetry will be
restored.

Let us recapitulate the results found in this Section.
The physical picture that emerges in our study of separated two stack
$D3$ brane background is qualitatively different from the case of
continuously distributed $D3$ branes. As expected,  due to the broken
conformal symmetry, the correlation functions of the boundary SYM theory
in its Coulomb phase have terms which vanish only in the conformal limit.
There is a deformation term  much like in the case of spherical $D3$ brane
shell and conformal like correction which is of order $\frac{1}{N}$.  This
correction does not appear in the case of uniformly distributed $D3$
branes.  Futhermore, due to interaction between KK modes, off diagonal
correlators are nonvanishing and are of order $\frac{1}{N}$.  All correction
terms vanish in the conformal limit.The term proportional to
$\frac{1}{N}$ is not a string loop correction as we have used the
classical supergravity limit.


\section{Deformation terms and the spectrum}

In the previous section it was shown that the correlation
functions on Coulomb branch have deformation terms.
Schematically, they have the following form:
\begin{equation}
D_{n,n}(\kappa)\sim (c_1)\b_{n,n}(\kappa)+
(c_2)\b_n(\kappa)\b_{n-2,n}(\kappa)+
(c_3)\b_{n}(\kappa)
\notag
\end{equation}
\begin{equation}
D_{n,n+2}(\kappa)\sim (c_4)\b_{n,n+2}(\kappa)+
(c_5)\b_{n,n}(\kappa)+
(c_6)\b_{n+2}\b_{n,n}(\kappa)
\notag
\end{equation}
where the coefficients $c_i$ can be read off from the explicit form of
$D_{n,m}$ and as before:
$$
\b_{n,m}(\kappa)=\b_{n,m}^{(1)}(\kappa)+\b_{n,m}^{(2)}(\kappa)
$$
In the above formulae we have separated $\b_{n,m}$ into
$\b_{n,m}^{(1)}$, the part
has in the denominator a product of a Bessel function and it
derivative, and $\b_{n,m}^{(2)}$ which contains in the denominator the
square of a Bessel function times its derivative.  We note also that
$\b_n$ contains in the denominator $I_{n+2}I_{n+2}'$.

In order to analyze the structure of $D_{n,m}$ we
perform a Wick rotation $\vec
k\to -i\vec k$. One can use the transformation properties of the modified
Bessel functions:
$$
I_\nu(-ik)=e^{-i\pi\frac{\nu}{2}}J_\nu(k)
$$
$$
K_\nu(-ik)=\frac{i\pi}{2}e^{i\pi\frac{\nu}{2}}
\left[J_\nu(k)+iN_\nu(k)\right]
$$
to obtain the relevant expression for the case of Minkowski space time.
Since the expressions are rather complicated we will analyze only the
denominators. From the above formulae, one can see that there are two
kinds of poles. Simple poles come from the zeroes of:
\begin{equation}
J_{n+2}(\kappa)J_{n+2}'(\kappa)=0;\quad\text{or}\quad
J_{n+2}(\kappa)J_{n+2}'J_{n}(\kappa)J_{n}'(\kappa)=0
\end{equation}
and poles of order two are determined by
$$
[J_{n+2}(\kappa)J_{n+2}'(\kappa)]^2=0.
$$

 We interpret these as describing the  spectrum
of glueballs with double degeneracy. It is reasonable to
expect that the two stacks might introduce some
degeneracy in the spectrum since they contribute identically to the
boundary.  The spacing between the different resonant frequences is
given in units of $\frac{L^2}{d}$.  It is interesting to note that we
obtain the same resonant frequencies as in the case of $D3$ brane
shell.  However the resonances are sharp.  One possible way to resolve
this puzzle, as proposed in \cite{GR} is to consider effect of
absorption by the branes which introduces finite width to these
resonances.  It would be very interesting to study the stability of
these resonances and their eventual decay into gauge bosons.

We show below that the deformation terms contain logarithmic
corrections to the two point correlators due to the breaking of
conformal symmetry. We will consider only the diagonal term in (64)
(because it
is leading order in $\frac {d}{L^2}$)
Then we find,
$$
D_{0,0}(\kappa)\sim (\frac {\kappa}{2})^2 \tilde\beta_{0,0}+\dots
$$
where ellipses stand for contributions from the mixing.  In the expression for
$\tilde\beta_{0,0}$ we perform a Wick rotation and
 we find
 $$\tilde\beta_{0,0}=-\frac{N_2J_2'+N_2'J_2}{J_2J_2'}=-\frac{N_2'}{J_2'}
-\frac{N_2}{J_2}
$$
Thus we have to evaluate the expression \footnote{Since we want to
demonstrate appearance of $log$ corrections we will not keep track of
numerical factors.}
$$\frac{1}{(2\pi)^4}\int d^4k e^{-i\vec k.(\vec x-\vec
y)}k^4\tilde\beta_{0,0}(\kappa)=\frac{1}{(2\pi)^2}
\frac{1}{|\vec x-\vec y|}\int dk k^6 J_1(k|\vec x-\vec
y|)\tilde\beta_{0,0}(\kappa)
$$
 Changing the variable $k\to k|\vec x-\vec y|$ one finds that
$$
\frac{1}{(2\pi)^4}\int d^4k e^{-i\vec k.(\vec x-\vec y)}
\tilde\beta_{0,0}(\kappa) = \frac{1}{(2\pi)^2} \frac{1}{|\vec x-\vec
y|^8}\int dk k^6 J_1(k)
\tilde\beta_{0,0}\left(\frac{d}{L^2}\cdot\frac{k}{|\vec x-\vec y|}
\right)
$$
In order to obtain the leading order  contribution, one can use the
series representation for the Bessel function $N_2$ in
$\tilde\beta_{0,0}$
$$
\pi N_2(\kappa|\vec x-\vec y|)=J_2(\kappa|\vec x-\vec y|)
log\frac{\kappa|\vec x-\vec y|}{2}+ \dots .
$$
In above the dots stand for terms of positive powers of $\kappa|\vec
x-\vec y|$ which for a local theory is small ($|\vec x-\vec y|$ is
finite, but infinitesimally small).  Let us consider first the term
$N_2/J_2$.  The integral becomes
$$
\frac{1}{(2\pi)^4}\int d^4k e^{-i\vec k.(\vec x-\vec y)}
\tilde\beta_{0,0} = \frac{1}{(2\pi)^2} \frac{1}{|\vec x-\vec
y|^8}\int dk k^6 J_1(k) \left( log\,k|\vec x-\vec y| + \dots\right)
$$
 To compare our results with known results we will restrict to energies
 below the mass of the $W$-boson.  In this case the integral over $k$
 is finite and the logarithmic correction to the two point correlation
 function is reproduced.  Analogous considerations hold in the case of
 $N_2'/J_2'$.  We note one can repeat these steps  for the non-diagonal
 part of $D_{0,0}$ which will produce a logarithmic correction to the
 mixed term.  In the above range of energies, from the gauge theory side,
 the physics is as follows \cite{Costa},[18].  Since the conformal
 symmetry is broken by Higgs scalar field, we will have a
 renormalization group flow from the conformal point of $SU(N)$ to IR
 fixed point with $SU(N/2)$.  If we consider the case of a D3 brane
 probe, the corresponding picture on the gravity side is as a string
 stretched between the D3 brane and the horizon of AdS and can be
 described by the Born-Infeld action which is a Yang-Mills theory plus
 higher derivative terms.  Then $d/\alpha'$ will correspond to the mass
 of a $W$-boson and the symmetry will be broken from $SU(N)$ to
 $S(U(N-1)\times U(1))$.  Now, one can proceed by adding more and more
 D3 branes to the single one until the resulting theory becomes
 $S(U(N/2)\times U(N/2))$.  All of this have correct gravity
 approximation when $g_sN>>1$, i.e.  large N.  In the large $N$ limit
 only planar diagrams contribute.  The logarithmic correction then has
 an interpretation as due to processes where a pair of virtual gauge
 particles is produced.  The ultraviolet cut off is related to the
 gravity mass gap and in the strong coupling limit it leads to a
 condensate of gauge particles.  It is shown in the above that our
 deformation term has poles in the Wick rotated $k$ plane that
 correspond to bound states of glueballs.

The main feature in our calculation is contained in the second term in
(64) as well as in the non-vanishing values for the mixed correlators,
i.e.  correlators of conformal operators of different conformal
weight.  Note that these terms are proportional to $\frac{d^2}{\alpha'
N^2}\sim \frac{m_W^2}{g_sN}$.  If we integrate out all the massive
particles in the gauge theory, we will obtain just the logarithmic
correction found in \cite{Costa} and reproduced in the above.  If,
however, we integrate out all the massive states above some
characteristic mass scale ($m_{charact}<< \frac{d}{L^2}$), one can
expect to have non-zero  correlation functions between conformal
operators of different conformal weight.  This is also the structure
of our mixed terms.  Moreover, string loop corrections coming from the
Born-Infeld action will be proportional to the above characteristic
mass and will be negligible compared to the $\frac{1}{N}$ correction  term
for the correlators.

\section{Conclusions}

In the present paper, we have studied the Coulomb branch of SYM theory
using AdS/CFT correspondence. Our considerations are based on the
simplest example of two stacks of $N/2$ parallel $D3$ branes separated
by a distance $2d$ which break  $SU(N)$ symmetry to
$S(U(N/2)\times U(N/2))$. Inspite of its apparent simplicity,
the problem has turned out to be complicated
but interesting. This complication mainly comes from the coupling between
different $n$-wave dilatonic KK modes in the equations of motion.
These equations are solved in the leading ($1/N$) order.
 We have shown that
the coupling between KK dilatonic modes  causes the appearance of
mixed correlation functions
which are of order $1/N$.
According to AdS/CFT correspondence
principle, the operators ${\mathcal O}^n$
correspond to $tr(F^2X^n)$.
It is
obvous from the explicit form of the correlation functions that in the
conformal limit ($d\to 0$, or Higgss vev tends to zero) the
$\frac{1}{N}$ correction term in the diagonal correlators and the mixed
correlation functions vanish.

All the correlation functions have also deformation terms.
The analysis in the
phase space, after a Wick rotation, shows that these functions have a
resonant structure - double poles at appropriate resonant frequencies.
We interpret these double poles as the degenerate spectrum of glueballs.

It will be interesting to study the correlation functions beyond this
level of approximation. We expect more conformal-like terms to
appear in the correlations functions.
We also have also to learn more about the spectrum of the glueballs
and to look for a mechanism of deriving more realistic masses.

\vspace*{.8cm}

{\bf Acknowledgements:}  R.R. would like to thank Simon Fraser
University for warm hospitality. This work has been supported by an
operating grant from the Natural Sciences and Engineering Research
Council of Canada.

\vspace*{.8cm}


\vspace*{1cm}

\section{Appendix: The particular solutions of (\ref{1fe>n}).}

In this Appendix we will find the particular solutions for the first
order equations of motion. We have, in fact, to find particular solutions
for three cases:
\begin{align}
f_n''(\hxi)+\frac{1}{\hxi}f_n'(\hxi)-\left(1+\frac{(n+2)^2}{\hxi^2}\right)
f_n(\hxi)=
\begin{cases}
&\hxi^2I_{n}(\hxi)\cr
&\hxi^2I_{n+2}(\hxi)\cr
&\hxi^2I_{n+4}(\hxi)
\end{cases}
\quad n=1,\dots
\label{part}
\end{align}
Let us consider the following expression:
$$
{\mathcal D}_nH\equiv H''+\frac{1}{\hxi}H'(\hxi)-\left(1+\frac{(n+2)^2}
{\hxi^2}\right)H(\hxi)
$$
Since the right hand side of eq. (\ref{part}) contain Bessel functions
times powers of $\hxi$ it is instructive to consider the function $H$ of
the form:
\begin{equation}
H=\hxi^aI_m(\hxi)
\label{ansa}
\end{equation}
After simple transformations and using the Bessel equation for $I_m$ we
find:
\begin{equation}
{\mathcal D}_nH=\hxi^{a-2}\left[2a\hxi
I_m'+\left(a^2+m^2-(n+2)^2\right)I_m\right]
\label{basic}
\end{equation}
If we substitute for $I_m'$ in the above equation:
\begin{equation}
\hxi I_m'=\hxi I_{m+1}+mI_m
\label{tr1}
\end{equation}
we find:
\begin{equation}
{\mathcal D}_nH=\hxi^{a-2}\left[2a\hxi
I_{m+1}+\left((a+m)^2-(n+2)^2\right)I_m\right]
\label{basic1}
\end{equation}
Let us choose the parameters $a$ and $m$ in the following way:
$$
a+m=n+2
$$
Therefore:
\begin{equation}
{\mathcal D}_nH=\hxi^{n-m+1}2(n-m+2)I_{m+1}(\hxi)
\label{basic2}
\end{equation}
For $m=n-1$ one find:
\begin{equation}
\left[
\frac{d^2}{d\hxi^2}+\frac{1}{\hxi}\frac{d}{d\hxi}
-\left(1+\frac{(n+2)^2}{\hxi^2}\right)\right]
\frac{\hxi^3 I_{n-1}(\hxi)}{6}=
\hxi^2 I_n(\hxi)
\label{solu1}
\end{equation}
which solves the first case in eq. (\ref{part}). For $m=n$ we have:
\begin{equation}
\left[
\frac{d^2}{d\hxi^2}+\frac{1}{\hxi}\frac{d}{d\hxi}
-\left(1+\frac{(n+2)^2}{\hxi^2}\right)\right]
\frac{\hxi^2 I_{n}(\hxi)}{4}=
\hxi I_{n+1}(\hxi)
\label{sol1}
\end{equation}
To solve  the second case one can use the properties of the Bessel
function $I_{n+2}$:
$$
\hxi I_{n+2}(\hxi) =\hxi I_{n}(\hxi)-2(n+2)I_{n+1}(\hxi)
$$
and therefore:
$$
\hxi^2 I_{n+2}(\hxi) =\hxi^2 I_{n}(\hxi)-2(n+2)\hxi I_{n+1}(\hxi)
$$
Using the above we have the solution for the second case.

To obtain the solution for the third case instead of eq. (\ref{tr1}) one
can use the following property of Bessel functions:
$$
\hxi I_{m}'(\hxi)=\hxi I_{m-1}(\hxi)-mI_{m}(\hxi)
$$
Then we will have:
\begin{equation}
{\mathcal D}_nH=\hxi^{a-2}\left[2a\hxi
I_{m-1}+\left((a-m)^2-(n+2)^2\right)I_m\right]
\label{basic3}
\end{equation}

Let us choose:
$$
a=m-n-2;\qquad m=n+5
$$
With this choice we find:
\begin{equation}
\left[
\frac{d^2}{d\hxi^2}+\frac{1}{\hxi}\frac{d}{d\hxi}
-\left(1+\frac{(n+2)^2}{\hxi^2}\right)\right]
\frac{\hxi^3 I_{n+5}(\hxi)}{6}=
\hxi^2 I_{n+4}(\hxi)
\label{soluti}
\end{equation}

Now, consider the system of equations  containing
the modified Bessel functions $K_n$:
\begin{align}
f_n''(\hxi)+\frac{1}{\hxi}f_n'(\hxi)-\left(1+\frac{(n+2)^2}{\hxi^2}\right)
f_n(\hxi)=
\begin{cases}
&\hxi^2K_{n}(\hxi)\cr
&\hxi^2K_{n+2}(\hxi)\cr
&\hxi^2K_{n+4}(\hxi)
\end{cases}
\quad n=1,\dots
\label{part2}
\end{align}
we find:
\begin{align}
&\left[
\frac{d^2}{d\hxi^2}+\frac{1}{\hxi}\frac{d}{d\hxi}
-\left(1+\frac{(n+2)^2}{\hxi^2}\right)\right]
\left(-\frac{\hxi^3 K_{n-1}(\hxi)}{6}\right)=
\hxi^2 K_{n}(\hxi)
\label{soluti1}
\\
&\left[
\frac{d^2}{d\hxi^2}+\frac{1}{\hxi}\frac{d}{d\hxi}
-\left(1+\frac{(n+2)^2}{\hxi^2}\right)\right]
\left(-\frac{\hxi^2 K_{n}(\hxi)}{4}\right)=
\hxi K_{n+1}(\hxi)
\label{soluti2}
\\
&\left[
\frac{d^2}{d\hxi^2}+\frac{1}{\hxi}\frac{d}{d\hxi}
-\left(1+\frac{(n+2)^2}{\hxi^2}\right)\right]
\left(-\frac{\hxi^3 K_{n+5}(\hxi)}{6}\right)=
\hxi^2 K_{n+4}(\hxi).
\label{soluti3}
\end{align}


\begin{thebibliography}{99}
\bibitem{Malda} J. Maldacena, "The Large N Limit of Superconformal Field
Theories and Supergravity", Adv. Theor. Math. Phys. 2
(1998) 231, hep-th/9711200.
\bibitem{GKP} S. S. Gubser,I. R. Klebanov
and A. M. Polyakov, Gauge Theory Correlators from Non-critical String
Theory, Phys. Lett. B428 (1998) 105, hep-th/9802109.
\bibitem{W} E. Witten, Anti De Sitter Space
and Holography, Adv. Theor. Math. Phys. 2 (1998) 253, hep-th/9802150.
\bibitem{KW} I.Klebanov and E.Witten, AdS/CFT Correspondence and
Symmetry Breaking, Nucl. Phys. {\bf B556} (1999 89, hep-th/9905104.
\bibitem{Aha} O.Aharovy, S.S.Gubser, J.Maldacena, H.Ooguri, Y.Oz,
Lagre $N$ Theories, String theory and Gravity, hep-th/9905111.
\bibitem{t'H} G.t'Hooft, Nucl. Phys. {\bf B72} (1974) 461.
\bibitem{Pol} A.M.Polyakov, Nucl. Phys. {\bf B 68} (Proc. Suppl.)
(1998), hep-th/9711002.
\bibitem{CR} I.Chepelev and R.Roiban, Phys. Lett. {\bf 462} (1999) 74
hep-th/9906224.
\bibitem{GR}S.Giddings and S.Ross, hep-th/9907204.
\bibitem{FGPW} D.Z.Freedman, S.S.Gubser, K.Pilch and N.P.Warner,
hep-th/9906194.
\bibitem{BSf} A.Brandhuber and K.Sfetsos, hep-th/9906201.
\bibitem{BKL} V.Balasubramanian, P.Kraus, A.Lawrence and S.Trivedi,
Phys. Rev. {\bf D59} (1999) 389, hep-th/9808017.
\bibitem{Sf} K.Sfetsos, JHEP 9901 (1999) 015, hep-th/9811167.
\bibitem{KLT} P.Kraus, F.Larsen and S.Trivedi,
JHEP 9903 (1999) 003, hep-th/9905171
\bibitem{Rang} S.Lee, S.Minwalla, M.Rangamani and N.Seiberg,
Adv. Theor. Math. Phys. 2 (1998) 697, hep-th/9806074.
\bibitem{Fr} O. De Wolfe and D.Freedman, hep-th/0002226.
\bibitem{KR} H.J.Kim, L.J.Romans and P.van Nieuwenhuizen, Phys. Rev.
{\bf D32} (1985) 389.
\bibitem{Costa} Miguel S. Costa, hep-th/9912073.
\bibitem{costa} Miguel S.  Costa, hep-th/0003289.
\end{thebibliography}
\end{document}